\documentstyle[12pt]{article}
\begin{document}
\newcommand{\be}{\begin{equation}}
\newcommand{\ee}{\end{equation}}
\newcommand{\bea}{\begin{eqnarray}}
\newcommand{\eea}{\end{eqnarray}}
\newcommand{\Dt}{\Delta}
\newcommand{\pf}{\Omega_{pl}}
\newcommand{\pff}{\Omega_{pl}^2}
\newcommand{\Om}{\Omega}
\newcommand{\omg}{\omega}
\newcommand{\epi}{\epsilon_{\infty}}
\newcommand{\eps}{\epsilon}
\newcommand{\Gm}{\Gamma}
\newcommand{\ga}{\gamma}
\newcommand{\lj}{\lambda_J}
\newcommand{\gm}{\gamma}
\begin{center}
{\Large\bf Plasma Oscillations in Layered Superconductors.}\\
\vspace{3mm}
Valery L. Pokrovsky\\
{\it Department of Physics, Texas A\&M University,\\
College Station, TX 77843-4242, USA\\
and\\
Landau Institute for Theoretical Physics,\\
Chernogolovka 142432, Russia.}
\end{center}
\section{Introduction.}
Very soon after publication of the famous BCS work explaining
the puzzle of the superconductivity, N.N. Bogolyubov and coworkers
\cite{bogolyubov} have proposed their version of the theory. One of
new results they have obtained was the discovery of a collective
mode - an oscillation of the Cooper pair density with the energy
smaller than $2\Dt$. P.W. Anderson has indicated that this
collective mode can not be found experimentally since the Coulomb
forces neglected in the above mentioned work shift its energy
to the plasma frequency, i.e., to the high-ultraviolet range in
which the superconductivity is inessential. The interest for plasmons
in superconductors revived a little in the search for the mechanism
of the High-$T_c$ superconductivity. However, a real surge of
interest to this problem occured after experimental observations
of the plasma edge in the reflectivity of High-$T_c$ superconductors
$La_{2-x}Sr_{x}CuO_4$ and $YBa_2Cu_3O_{8-y}$ \cite{koch},
\cite{tamasaku}. It is worthwile to mention that a theoretical
prediction have preceded the experiment \cite{mishonov}. For
external reasons, the paper \cite{mishonov} has
been published only a long time after its completion.
\par
Here we present a brief review of the experiments and theoretical
developments in the field. The theoretical works will be
presented in more detail, given the author's specialization.
I hope that this imbalance will be compensated by the papers
by O.K.C. Tsui, N.P. Ong and J.B. Peterson, and by P.M. Mueller.
\par
The content of the review is as follows. In the second section
we review the most general relationships for plasma frequency (PF)
and dielectric function of the electron plasma in normal metals
and their modification for layered normal metals. In the third
section we review the experimental observations of the plasma
edge in High-$T_c$ superconductors (HTSC). In the fourth section
a simple two-fluid model of the plasma oscillations is presented.
The fifth section is devoted to the BCS theory of the same
phenomenon. Peculiarities of Josephson-linked layered superconductors
are considered in Section 6. The experimental observations
of the resonance plasma attenuation are discussed in Section 7.
Mechanisms of dissipation in the Josephson-linked
superconductors are discussed in Section 8. A novel effect
of selectional transparency is discussed in Section 9.
The conclusions are collected in Section 10.\\
We employ abbreviations SC for superconductors and OP
for the order parameter.

\section{Plasma resonance in normal metals.}
The plasma oscillation is one of the simplest and most fundamental
manifestations of the Coulomb interaction. Due to the long-range
character of the Coulomb forces, its spectrum starts with a rather
high frequency given by the well-known formula:
\be
\pff = \frac{4\pi ne^2}{m\epi}
\label{ompl-gen}
\ee
where $n$ and $m$ are the electron density and mass respectively,
$\epi$ is the dielectric constant contributed by atomic and ion
reminder shells. For usual values in solids $n\sim 10^{22}\div
10^{23}cm^{-3}$, $m=10^{-27}g$ and $\epi\sim 1$, the PF
$\pf\sim 10^{15}\div 10^{16}s^{-1}$ lies in the high-ultraviolet
range. The variation of $n$ and $m$ in metals are not large.
Therefore the characteristic value of the PF is always
rather high. This is the reason why plasma oscillations do not
play an important role, either in the thermodynamics or in
transport phenomena in conventional metals.
\par
The situation changes in layered metals, if the electron propagation
perpendicular to the layers (the $c$-axis) is mediated by
tunneling processes. Applying the tight-binding approximation
for the $c$-direction one finds a following dispersion law for
electrons:
\be
\eps ({\bf p}, p_z)\,=\,\eps_{\parallel}({\bf p})+t(1-\cos\frac{p_zd}
{\hbar})
\label{dis}
\ee
where ${\bf p}, p_z$ are in-plane and perpendicular-to-plane
components of the momentum, $t$ is the overlap integral
and $\eps_{\parallel}(\bf p)$ is the in-plane dispersion. Assuming
it to be $\eps_{\parallel}({\bf p})={\bf p}^2/2m$ for simplicity, we
find that for $t\ll \eps_F$ the Fermi-surface is a slightly
corrugated circular cylinder. Solving the kinetic equation
with such a Fermi-surface, we find that Eq. (\ref{ompl-gen})
holds for the electric field in-plane, but for the
perpendicular-to-plane electric field the PF is
much smaller \cite{PP}:
\be
\pff\,=\,\frac{4\pi n_e^2}{\epi M}\frac{m}{M}\left(
\frac{\hbar}{p_Fd}\right)^2
\label{ompl-perp}
\ee
where $M=\hbar^2/td^2\gg m$ and $d$ is the lattice constant in
the $c$-direction. Note that the ratio $\Om_c/\Om_{ab}\propto
m/M$ instead of $\sqrt{m/M}$ as it could be naively expected and
what is correct for an elongated ellipsoid. The reason is
the change of topology: the cylinder is an open surface. This
simple observation allows one to understand a drastic reduction
of the PF which has been observed experimentally
\cite{koch}, \cite{tamasaku}: $\pf$ for $LaSrCuO$ and for $YBCO$
is in the far infrared region. This agrees well with a comparatively
low carriers densities found in the Hall-effect measurements
and high ratio $M/m\sim 100\div 2,500$ found in experiments
with the tork produced by the vortex lattices \cite{farrel}.
The anisotropy parameter in Bi 2:2:1:2 is even larger
\cite{Yye}, $M/m\sim 40,000$. Therefore we predicted that plasma
frequency in this compound lies in the microvawe region \cite{PP}.
This theoretical prediction has been confirmed in the experiment
\cite{tsui1}, though the experimenters were not aware of the
theoretical prediction.
\par
Unless the collision time for quasiparticles $\tau$ is small
enough, there is no substantial difference between the plasma
oscillations in normal metals and superconductors. In both
cases they are the charge density oscillations with the
restoring force determined by the electrostatics. The situation
is very different for small collision time $\pf\tau\ll 1$.
Then normal carriers almost do not participate in the plasma
oscillations. We return to the analysis of this problem later.
\par
The PF may be alternatively determined as a frequency
at which the dielectric function $\eps (\omg )$ is zero. For the
collisionless plasma, the dielectric function is:
\be
\eps (\omg)\,=\,\epi \left( 1-\frac{\pff}{\omg^2}\right)
\label{eps-0tau}
\ee
The reflectivity for an EM wave is determined by the Fresnel
formula:
\be
R\,=\,\left|\frac{\sqrt{\eps (\omg )}-1}{\sqrt{\eps (\omg )}+1}
\right|^2
\label{fresnel}
\ee
According to Eqs. (\ref{eps-0tau}, \ref{fresnel}) $R=1$ at
$\omg <\pf$ and $R<1$ at $\omg >\pf$. This effect is called
the plasma edge. Its sharpness depends on the value $\epi$.
For $\epi >1$, the graph of $\eps (\omg )$ has a characteristic
beak (Fig 1). Introducing a phenomenological collision time
$\tau$ and the electron attenuation $\Gm =1/\tau$, one
finds instead of (\ref{eps-0tau}):
\be
\eps (\omg)\,=\,\epi \left( 1-\frac{\pff}{\omg (\omg+i\Gm})\right).
\label{eps-tau}
\ee
The plasma edge is smeared out at large $\Gm\gg \pf$ or small
$\tau$. Theoretical curves of $R(\omg )$ are shown on Fig.1.

\section{Experimental observation of the plasma edge in
superconductors.}
The first observation of the plasma edge has been reported by
Koch et al. \cite{koch} in 1990. They used a single crystal of
$YBCO$ specially grown in such a way that the layers were
oriented perpendicularly to the surface of a sample and the $c$-axis
was directed parallel to the surface. The electric vector in the
incident EM wave was directed along the the $c$-axis. This geometry
enabled the authors to observe the plasma edge.
They have found a sharp decrease of the reflectivity at a
frequency $\omg \approx 12 meV$ in the superconducting state ($T< 90K$)
and very smooth decrease of reflectivity in the normal state
(Fig. 2).
Unfortunately, the plasma edge in $YBCO$ occured to be close to
other features of the spectrum associated with optic phonons,
thus interfering with a thorough study of plasma oscillations.
\par
Tamasaku et al. \cite{tamasaku} have used very good single
crystals of $La_{2-x}Sr_{x}CuO_4$ in the same geometry
which enabled the authors to observe the plasma edge reliably (Fig. 3).
They also did not observe any edge effect in the
normal state and a rather sharp edge already at a temperature just
slightly below the transition temperature. A systematic decrease
of the PF with the temperature has been found. Their measurements
are compatible with the assumption that $\pff\propto\Dt^2$.
\par
Two questions naturally arise in connection with these
measurements.\\
1. Why there is no plasma edge in the normal state while it
occurs immediately in the superconducting state?\\
2. Why PF depends strongly on the temperature?\\
At a qualitative level these questions can be answered without detailed
calculations \cite{PP}, \cite{marel}. The absence of the plasma edge in the
normal state can be explained by strong collisions $\pf\tau\ll 1$.
In the superconducting state a part of electrons is confined in
Cooper pairs (superconducting carriers) which are not scattered.
It explains why the plasma oscillations occur in the
superconducting state. The same idea explains the temperature
dependence of the PF: not the full, but only
the superfluid density $n_s$ enters Eqs. (\ref{ompl-gen},
\ref{ompl-perp}). The superfluid density in an anisotropic
SC is a second-rank tensor. The superfluid
density varies from the total electron density at $T=0$ in a
clean SC till zero at $T=T_c$.

\section{Two-fluid theory of the plasma resonance.}
The two-fluid model of SC by Gorter and Kasimir
(see for example \cite{rickayzen}) is a simplest
phenomenological way to understand the
plasma oscillations in SC. It has been used
in fact by T. Mishonov \cite{mishonov}, and also in a great extent 
by D. van der Marel et al. \cite{deMarel} and later by Tachiki et al.
\cite{tachiki}. In the framework of this model, the dielectric
function is written in the following form:
\be
\eps (\omg )\,-\,\epi\left( 1-\frac{\Om_s^2}{\omg^2}-
\frac{\Om_n^2}{\omg (\omg +i\Gm )}\right).
\label{2-fluid}
\ee
The terms with $\Om_s^2$ and $\Om_n^2$ represent contributions
of the superfluid and normal parts of the electron liquid,
respectively. As it was emphasized in Section 3, the superfluid
part of the liquid is not subject to any dissipation whereas the
normal part contains the conventional dissipative coefficient
$\Gm$. If $\pf\tau\gg 1$, then $\pff =\Om_s^2+\Om_n^2$, i.e.
all carriers participate in the plasma oscillation. In the
opposite case $\pf\tau\ll 1$, one finds $\pf =\Om_s$, in agreement
with what was expected. Note that in the latter case the imaginary
part of $\eps$ is small:
\be
{\cal I}m\,\eps\,=\,\frac{\epi\Om_n^2\tau}{\Om_s};\,\,\,\,\,\,\,
(\omg\approx\Om_s).
\label{im-eps}
\ee
Thus, the stronger are collisions the weaker is the dissipation
of the plasma oscillations.
This paradoxical result is explained by different roles played
by normal carriers at different $\pf\tau$. If this parameter
is large, the normal carriers create the coherent plasma wave on
the same footing
as the superconducting electrons. If $\pf\tau$ is small, the
coherent motion is created by superconducting electrons only
whereas the normal electrons, being partly involved into this
motion, dissipate the energy. The sooner they are put in equilibrium
by collisions the more adiabatically they follow an instant
value of the OP, and the smaller are losses.
\par
Though the two-fluid model describes qualitatively the plasma
oscillations it has several important failures. First, the
very notions of the superconducting and normal densities 
$n_s$ and $n_n$ can be defined reasonably only for a rather 
clean SC. These definitions presume the conservation
of the momentum which is violated by impurities, phonons etc.
Second, the phenomenological definition of the
collision time does not enable one to calculate its temperature
and frequency dependence. Third, the two-fluid model does
not take into account the symmetry of the superconducting
OP making no difference between $s$- and $d$-pairing.

\section{BCS theory of plasma oscillations.}
The BCS theory of plasma oscillations has been proposed in
\cite{PP} to overcome the failures of the two-fluid model.
It is a microscopic theory based on the BCS Hamiltonian
with the dispersion law given by Eq. (\ref{dis}). The
OP in the unperturbed SC is
assumed to be uniform in the real space and very anisotropic
in the relative momentum space. Only the elastic scattering
by impurities has been considered. The scatterers were assumed
to change the electron momentum in plane, but to be unable to
transfer an electron from one plane to an ajacent plane. It
means that the scattering amplitude does not depend on the
z-component $p_z$ of the momentum. One more simplification is
provided by an assumption of the full in-plane isotropy.
The problem is to calculate the Pippard kernel $Q(\omg , \bf q)$.
The Pippard kernel relates the Fourier components of the
current density $\bf j$ and the vector potential $\bf A$:
\be
{\bf j}(\omg ,{\bf q})\,=\,-Q(\omg ,{\bf q}) {\bf A}(\omg ,{\bf q})
\label{pippard}
\ee
The relationship (\ref{pippard}) is written for a special gauge
$\nabla{\bf A}=0$. Once the Pippard kernel is known the dielectric
function can be calculated as:
\be
\eps (\omg, {\bf q})\,=\,\epi-\frac{4\pi c}{\omg^2}Q(\omg ,\bf q)
\label{eps-pip}
\ee
In the real space the relationship (\ref{pippard}) reads:
\be
j_{\alpha}(\omg ,{\bf r})=-\int Q_{\alpha ,\beta}
(\omg ,{\bf r, r}^{\prime}) A_{\beta}(\omg ,{\bf r}^{\prime})d^3x
\label{pipp-real}
\ee
The relationship (\ref{pipp-real}) is valid at a fixed configuration
of impurities which violates the translational invariance. According
to the Kubo formula, the response function (Pippard kernel) is
expressed in terms of the retarded current-current Green function
which in turn can be obtained as an analytical continuation of the
Matsubara Green function:
\be
Q_{\alpha ,\beta}^{M}(\omg_n ,{\bf r, r}^{\prime})=
\frac{iT}{c}\int_0^{1/T}d\tau e^{i\omg_n\tau}\langle
j_{\alpha}(\tau ,{\bf r})j_{\beta}(0, {\bf r}^{\prime})
\rangle
\label{matsubara}
\ee
with $\omg_n=2\pi nT$ and integer $n$. The retarded response is
the analytical continuation of this function from discrete
points $\omg=i\omg_n$ in the upper half-plane of the complex
plane $\omg$. The next averaging over the random location
of impurities should be performed in the end.
\par
Abrikosov and Gor'kov \cite{AG} have developed a special
diagrammatic technic for this averaging. In a simplest situation
of weak scatterers, the problem is reduced to the solution of
an analog of the Boltzmann kinetic equation for SC
\cite{PS}, \cite{el}. An essential complication for the
superconducting case is the existence of several distribution
functions instead of one in the normal case. Not only the
occupation numbers of quasiparticles, but also the condensate
wave function varies locally. The solution of the kinetic
equation is generally nontrivial even in the normal case.
Fortunately, our special geometry together with the fact
that the penetration depth is much larger than any other
geometrical scale in the problem makes this problem exactly
solvable. More accurately, it can be solved in the leading
order over small tunneling amplitude $t$ if one employs the
independence of the impurity scattering amplitude on $p_z$.
For the most interesting $zz$-component of the Pippard
kernel the result is:
\be
Q_{zz}(\Om ,T)\,=\,\frac{e^2t^2md}{4\pi c\hbar^4}K(\omg ,T)
\label{Q-K}
\ee
\bea
K(\omg ,T)=\int_{-\infty}^{\infty}\frac{d\eta}{4}
\left[ \left( 1+\frac{\eta_+\eta +\Dt^2}{\eps_R(\eta_+)\eps_A(\eta )}
\right)\frac{\tanh (\eta_+/2T)-\tanh (\eta /2T)}
{\eps_R(\eta_+)-\eps_A(\eta) +i/\tau}\right.
\nonumber\\
\left. +\tanh\frac{\eta}{2T}\left(\frac{1-\frac{\eta_+\eta +\Dt^2}
{\eps_R(\eta_+)\eps_R(\eta)}}
{\eps_R(\eta_+)+\eps_R(\eta) +i/\tau}+
\frac{1-\frac{\eta_-\eta +\Dt^2}{\eps_A(\eta_-)\eps_A(\eta)}}
{\eps_A(\eta_-)+\eps_A(\eta) -i/\tau}
\right)\right]
\label{K}
\eea
where $\eta_{\pm}=\eta\pm\omg$, $\eps_{R,A}=\sqrt{(\eta\pm i\delta)^2
-\Dt^2}$, the branch of the square root is chosen to be positive
at $\eta >\Dt$. The reader is referred to the original work \cite{PP}
for details.
\par
One of the most important results in the BCS theory of the plasma
oscillations is
the vanishing of the dissipation in a pure SC at
any $T<T_c$. Two conventional dissipation processes are the
Cherenkov absorption of light by quasiparticles and the Cooper
pair breaking, i.e., the creation of a pair of quasiparticles
with the absorption of a photon. The first process is
kinematically forbidden since the light velocity $c/\epi$ is
much larger than the Fermi velocity. The second process, specific
for SC, is kinematically allowed at $\omg >2\Dt$.
However, it is forbidden dynamically. The amplitude
of the quasiparticle-photon interaction is proportional to the
so-called Bogolyubov coherence factor \cite{rickayzen}
$(u_{\bf p+q}v_{\bf p}-v_{\bf p+q}u_{\bf p})^2$ where $\bf p$
is the electron momentum and $\bf q$ is the photon pomentum.
For the extreme of the normal skin-effect $q\rightarrow 0$
this factor is zero. Both processes are allowed in the
impure SC in which the distribution over
momentum at a fixed energy looses its delta-like character
and becomes a Lorenzian one. This argument explains why
the plasma edge is observable in the HTSC
which have been experimentally proven to have lines of nodes of
the OP and zero energy gap in the spectrum
\cite{wollman}, \cite{mathai}, \cite{ott}, \cite{dynes},
\cite{schrieffer}.
\par
Eqs. (\ref{eps-pip}, \ref{K}) show that what is called
$\Om_s$ and $\Om_n$ in the two-fluid model in the microscopic
BCS theory are complicated functions of all variables.
Even at very low temperature the analog of the superfluid
density $n_s$ depends rather sharply on $\tau$ varying from the
value of total electron density at $\tau\Dt\gg 1$ to a
value less by a factor $\sqrt{\tau\Dt}$ at $\tau\Dt\ll 1$.
\par
If $\pf\ll\Dt$, the PF is expressed in terms of the penetration depth:
$\pf =c/\lambda_c$. This is just the case for Bi 2:2:1:2. However,
in YBCO and LaSrCuO, $\pf$ is comparable with $\Dt$. In this
case, the PF can be found as a solution of a following
equation:
\be
{\cal R}e\,\eps (\pf )=0
\label{plfr-exact}
\ee
For this situation the exact result (\ref{K}) is especially
important.
\par
In a clean SC, ${\cal I}m Q=0$, as it has been shown earlier.
Therefore, $K=1$ at $\tau=\infty$, independently on
temperature and frequency. This is a generalization of the
relationship $\pff =\Om_s^2+\Om_n^2$ at $\Gamma =0$ in the
two-fluid model. The
transition between the static value $K\propto n_s(T)$ and
the pure limit $K=1$ occurs in a region of frequency
$\omg\sim 1/\tau$, very narrow in clean SC.
\par
Comparing BCS theory and the two-fluid theory of EM response
in layered SC we find that quantitatively the
two-fluid model is rather unsatisfactory. It is especially poor in
evaluation of the dissipation for which its error (a deviation
from the BCS theory) may be 1000\% and more. The reason is
that there is only one scale of frequency $1/\tau$ for
dissipation in the two-fluid model whereas the BCS theory
displays several equally important scales: $\Dt, T, 1/\tau ,
\Dt^2\tau, (\Dt\tau^2)^{-1}$.
\par
Though the BCS theory is a step forward in our understanding of
the phenomenon its premises are oversimplified. The most important of
these simplifications are as follows:\\
i) The BCS theory is a theory of weakly interacting electrons.
It may not work well for HTSC. Unfortunately the
existing theories of strongly correlated fermions do not reach
a level wich allows for a reliable calculation of the EM response.\\
ii) The scattering in the BCS model has been assumed to be
elastic scattering by weak impurity scatterers. It probably
dominates in the LSCO which becomes a SC at a sufficiently
high doping only. However, measurements of the Hall angle in
YBCO at low temperatures \cite{harris} have shown a high
purity of this compound. In combination with the plasma edge
observation \cite{koch} this leads to the conclusion that the
scattering rate is large near the transition temperature
($\pf\tau\ll 1$) and drops fast with decreasing the temperature.
\par
The problem of the scattering mechanism stays in a close connection
to the problem of the strong electron interaction. The famous linear-
in-temperature resistivity \cite{lin-resist} can hardly be explained
in the frameworks of the Fermi-liquid theory. For our purpose
we need the $c$-axis resistivity which behavior resembles 
semiconductors more than the \cite{uchida}. We can
argue that the variable-distance hopping processes determining
the static $c$-axis resistivity are irrelevant at plasma
frequency since they require a long time. Nevertheless,
the direct theoretical approach is much desirable.\\
iii) Our version of the BCS theory does not take into account a
strong  spatial modulation of the superconducting OP.
However, this is rather important for explanation of a strong influence
produced by the magnetic field on the PF. This
question is discussed in the next section.\\
iv) We have already mentioned that the experiments indicate a
nontrivial character of the pairing in HTSC with the nodes
of the OP. The BCS theory should be reformulated
for this situatuion.

\section{Plasma oscillations in Josephson-linked SC.}
In real HTSC the OP is deeply modulated in the $c$-
direction with the lattice periodicity. The coherence length in
the $c$-direction, $\xi_c$, for Bi, Tl and Hg based SC
is much less than the lattice constant $d$ in the same direction
practically at any $T<T_c$. For YBCO, $\xi_c$ becomes smaller than
$d$ at $T\approx 75K$ while $T_c=90K$. The OP in deeply modulated
superconductors is concentrated in thin layers near conducting
planes. The coupling between layers is a weak Josephson link.
The modulus $|\psi_n|$ of the OP in $n$-th plane is a  rigid
variable wich fluctuates only
weakly except very close to the transition temperature,
whereas the the phases $\varphi_n$ are Goldstone variables,
which are strongly fluctuating and are easily influenced 
by external fields.
In particular, the phase variables are drastically influenced
by the magnetic field. This is a general mechanism for a strong
dependence of the PF on the magnetic field.
\par
Specifically, the magnetic field penetrates into  a layered
SC as an
Abrikosov vortex lattice. We start with a geometry in which
the magnetic induction is
parallel to the layers. Then the vortices form a 2-dimensional
lattice with central lines situated just in the middle between
two ajacent planes. These vortices are similar to the Josephson
vortices occuring in a Josephson junction \cite{barone}.
Each of them creates an additional phase difference between
ajacent layers $2\pi$ on a characteristic scale $\lj=\gm d$,
the so called Josephson screening length (JSL),
where $\gm =\sqrt{M/m}=\xi_{ab}/\xi_c$ is the anisotropy coefficient.
The latter is about $30\div 50$ for YBCO  and not less than
200 for Bi 2:2:1:2, as it was established by several independent
experiments. Thus the JSL for BiSCCO reaches a macroscopic
value $\lj\sim 3000\AA$. If the magnetic field is so strong
that the vortices strongly overlap, i.e., the distance between
their centers is much less than $\lj$, the phase in the plane grows
almost linearly with the distance \cite{bul-clem}. In the
opposite limiting case of weakly overlapping vortices, the
phase difference $\Dt\varphi_{n, n+1}$ between $n$-th and
$n+1$-th layers grows rapidly by $2\pi$ on the scale of JSL
near each vortex and remains a constant until the next vortex line.
The boundary between these two regimes is the value of
magnetic field $B_J=\phi_0/\gm d^2$.
\par
To understand how the magnetic field influences the PF
we first establish an important relationship between
the PF and the maximal Josephson current $j_{max}$.
The analog of the
Pippard kernel $Q_{zz}$ or $n_s$ in the Josephson linked SC
is a value proportional to the maximum Josephson current
$j_{max}$. Indeed from the Josephson relationship:
\be
j_z=j_{max}\sin (\Dt\varphi_{n, n+1}-\frac{2A_zd}{\phi_0})
\label{joseph}
\ee
in a special gauge $\Dt\varphi_{n, n+1}=0$  and at small
$A_z$ we discover the linear dependence:
\be
j_z\,=\,-Q_{zz}A_z;\,\,\,\,\,\,\,\,Q_{zz}=\frac{2j_{max}d}{\phi_0}.
\label{j-A-lin}
\ee
By analogy with Eq. (\ref{eps-pip}) we find:
\be
\pff\,=\,\frac{8\pi cj_{max}d}{\epi\phi_0}
\label{PF-jmax}
\ee
The Josephson current can be found by differentiating
the corresponding free energy :
\be
F_J=\frac{j_{max}\phi_0}{2cd}\sum_n\int d^2x
\left[ 1-\cos\left(\Dt\varphi_{n,n+1} -
\frac{2A_zd}{\phi_0}
\right)\right].
\label{free}
\ee
A regular lattice of vortices stretched along the $y$-axis
causes a regular variation of $\Dt\varphi_{n, n+1}$
with $x$. We consider first $B\gg B_J$. Then
\be
\Dt\varphi_{n, n+1}\approx 2\pi\sqrt{\frac{B}{\gm\phi_0}}x
+\frac{B_J}{B}\sin \left( 2\pi\sqrt{\frac{B}{\gm\phi_0}}x
\right)
\label{phase-lin}
\ee
After averaging of Eq. (\ref{free}) over $x$ with the
$\Dt\varphi_{n, n+1}$ given by Eq. (\ref{phase-lin}) we find
\be
j_{eff}\,=\,j_{max}\frac{B_J}{2B}
\label{j-eff}
\ee
where $j_{eff}$ is the effective maximal current after the
averaging. In the opposite limiting case $B\ll B_J$ the
in-plane distance between vortices is $\sqrt{\gm\phi_0/B}$,
the out-of-plane distance between vortices is
$\sqrt{\phi_0/\gm B}$ and the range of essential variation
of phase near each vortex line is $d\lj$. Therefore
\be
j_{eff}=j_{max}(1-0(B/B_J)).
\label{j-small-B}
\ee
One should expect a sharp dependence of the PF
on the direction of the magnetic field. Such a dependence
has been predicted in \cite{bul-safar}. In particular the
authors of the work \cite{bul-safar} have predicted a reentrant
behavior of the PF vs the angle similar to the reentrant
behavior of the ac magnetic susceptibility and resistivity
discovered earlier \cite{mansky}, \cite{kadowaki}.
\par
Next we consider the magnetic field directed perpendicularly to
the layers. Each vortex in a layered SC consists of separate
plane vortices -- pancakes. If there is no irregularitiy in the
location of pancakes, the phase difference $\Dt\varphi_{n, n+1}$
created by them is zero and the magnetic field does not affect the
PF until $B\ll H_{c2}$. However, any disorder in
the pancake arrangement creates a random phase difference and
reduces efffectively $j_{max}$ and $\pf$. We consider the case of
the strong magnetic field when many pancakes enter a circle of
the radius $\lj$ \cite{BPM}:
\be
B\gg B_{J\perp}=\frac{\phi_0}{\lj^2}=\frac{\phi_0}{\gm^2d^2}.
\label{B_Jperp}
\ee
The condition (\ref{B_Jperp}) is much less restrictive than the
analogous condition for the parallel field: $B_{J\perp}\approx
200$ Gs for Bi 2:2:1:2. If the condition (\ref{B_Jperp}) is
satisfied, the phase $\Dt\varphi_{n, n+1}$ is effectively a sum
of a large number $N=B/B_{J\perp}$ of random terms. Therefore it obeys
the Gaussian statistics. The effective maximal current is
proportional to the Debye-Waller factor:
\be
j_{eff}\,=\,j_{max}\exp\left( -\frac{1}{2}\langle
(\Dt\varphi_{n, n+1})^2\rangle\right).
\label{debye}
\ee
The value $\Dt\varphi_{n, n+1}$ in our case is:
\bea
\Dt\varphi_{n, n+1}({\bf r})\,=\,\sum_{\nu}
\left[\Phi ({\bf r-r}_{n, \nu}) - \Phi ({\bf r-r}_{n+1, \nu})
\right]\nonumber\\
=\,\int\left(\rho_n({\bf r}) - \rho_{n+1}({\bf r})\right)
\Phi ({\bf r})d^2x
\label{phase-vort}
\eea
where $\Phi ({\bf r})=\arctan (y/x)$ and ${\bf r}_{n, \nu}$
is the vector coordinates of the center of a $\nu$-th
vortex in the $n$-th plane; $\rho_n({\bf r})=\sum_{\nu}
\delta ({\bf r-r}_{n, \nu})$ is the vortex density in the
$n$-th plane. After the averaging we find:
\be
\langle\Dt\varphi_{n, n+1}^2\rangle\,=\,2\mu\ln(\lj /a);
\,\,\,\,\,\,\,\,\mu=\frac{\pi}{4}\int r^2 {\cal K}({\bf r})d^2x
\label{exponent}
\ee
where ${\cal K}(\bf r)$ is a correlator:
\be
{\cal K}({\bf r})=\langle\rho_n({\bf r})\rho_n(0)\rangle-
\langle\rho_n({\bf r})\rho_{n+1}(0)\rangle
\label{kor}
\ee
If there is no correlation between ajacent layers, the correlator
${\cal K}(\bf r)$ turns into the correlator of a two-dimensional pancake
liquid. Thus, the measurement of the field dependence of the
PF provides an important information on the statistical properties
of the pancake liquid. Assuming the disorder within a plane
to be so strong that the only characteristic scale for ${\cal K}(\bf r)$
is the average distance between pancakes $a$, we find that the
dimensionless value $\mu$ is a numerical constant. As a consequence
we find a power law decrease of the PF with the
magnetic field:
\be
\pff\,=\, \Omega_{pl0}^2(B_{J\perp}/B)^{\mu}
\label{PF-B}
\ee
A reason for a strong in-plane disorder may be either thermal
fluctuations in the pancake liquid or a rather strong pinning in
a glassy phase. The first case corresponds to a temperature
above the irreversibility line, the second one -- below this
line. In a simplest irreversible situation a large number of
the strong pinning centers provide a random location of the
pancakes. A metastable state depends on the way of its
preparation. If the magnetic field is frozen, the vortices
find closest pinning centers and do not deviate much, at
least locally, from the regular Abrikosov lattice. If the field
switches on after the cooling, the vortices enter consequently
at the field is increased. The resulting vortex arrangement may
be highly chaotic.
\par
Consider a situation when there is no correlation between
positions of pancakes in ajacent planes, but the in-plane
crystal order persists over a distance $R$ much larger than
the in-plane lattice constant $a$. We also assume that $R\ll\lj$.
Then the exponent in Eq. (\ref{PF-B}) (see also Eq.(\ref{exponent})) 
becomes very sensitive
to the model of disorder and may be either much larger
or much smaller than one, as well as of the order of one. 
As a consequence,
the reduction of the PF may be much stronger than that caused
by a strong in-plane disorder. This striking result is
explained by the accumulation of a random interplane phase
shift on a large area $\propto R^2$.
\par
Thus the theory predicts a power-law decrease of the PF at
$B>B_{J\perp}$ and a strong in-plane disorder with the exponent
of the order of unity. The same result may be correct 
in the absence of the interplane correlation, but I can not exclude
a possibility of the strong reduction of the PF in this case. 
Finally, one should expect muvh weaker magnetic field effect, if
the interplane correlation is strong.
\par
The description of the Josephson-linked SC accepted in this section
is a kind of a long-wave-length phenomenology basically the same as
in the Lawrence-Doniach model \cite{doniach}. A microscopic
approach has been proposed by Artemenko and Kobel'kov \cite{artemenko}.

\section{Experimental observation of the resonance attenuation
at the PF.}
The first observation of the resonance attenuation has been
reported by O.K. Tsui et al. \cite{tsui1}. A sample of
$Bi_2Sr_2CaCu_2O_{8+\delta}$ (Bi 2:2:1:2) has been placed
into a waveguide. The sample was grown in the conventional way
with layers parallel to its surface. The resonance has been
observed in a range of frequencies between $30$ and $50 GHz$
and magnetic fields from 1 to 7T.
The absorption was recorded by a resonance bolometric
technique. During each experiment, the frequency was fixed
and the magnetic field varied. Rather sharp lines of
attenuation have been observed in the temperature interval
between 2.5 and 20K. The effect has been initially
ascribed to the cyclotron resonance. However, an additional
study has shown that it occurs only if the electric field has
a non-zero $c$-component, and tha the resonance frequency 
decreases with the
magnetic field \cite{tsui2}, \cite{matsuda}. This fact can be
considered as a convincing evidence in favor of the plasma-resonance 
origin of the phenomenon.
\par
Accurate measurements of the resonance \cite{tsui2} may be
summarized by an empirical formula valid at $T$ below the
irreversibility line:
\be
\pff (B,T)\,=\,AB^{-\mu}\exp (T/T_0)
\label{empiric}
\ee
where $A$ is a constant, $\mu\approx 0.8\div 1.0$ and $T_0\approx
12.5K$. In Section 6 we have derived the power low for $\pff$
at a condition of
strong in-plane disorder. Surprisingly the PF grows with the
temperature whereas according to a naive idea it should decrease
as the superfluid density. This growth can be qualitatively
explained in terms of partial thermal depinning of the vortices.
Depinned vortices tend to restore the positional order destroyed
by the pinning. Quantitatively we were able to prove that the
linear in $T$ correction to the PF is positive \cite{BPM}.
\par
In Bi 2:2:1:2 the contribution of quasiparticles to the PF is
probably always small. The reason is that they are strongly
damped at $T$ close to $T_c$ whereas the normal density is
very small at $T\ll T_c$. Besides of that $\pf\ll\Dt$ in
Bi 2:2:1:2.

\section{Dissipation mechanisms in Josephson-linked
SC.}
The resonance attenuation at $\omg =\pf$ requires an explanation. A most
obvious effect is the amplification of the electric field inside the
SC. In the experimental geometry the sample surface was
parallel to the layers, the electric field in the wave-guide was
perpendicular to the surface and the layers. From the continuity of 
the component of the electric
induction $D_z$ normal to the surface, it follows that
$E_{int}=E_{ext}/\eps (\omg)$,
where $E_{int}$ is the field inside the sample and
$E_{ext}$ is the field in the wave-guide. Since $\eps (\omg )$ is
almost zero at $\omg =\pf$, the electric field is strongly amplified
inside the sample. It creates a necessary premise for the resonance
attenuation, but this is not sufficient. The problem is that the electric
field in the $z$-direction creates the ac Josephson current in the
same direction, which is not subject to dissipation. A mechanism
transforming the electric field perpendicular to layers into 
the parallel one and the dissipation mechanism for the parallel
component of the electric field or current should be found.
Though this question is not fully solved we present here preliminary
ideas due to L. Bulaevsky and S. Pokrovsky 
without detailed calculations.
\par
Strongly pinned randomly located pancakes create large fluctuations
of the dielectric constant $\eps_{zz}$. The amplitude of this
fluctuations can be estimated as $1/\langle\cos\Delta\varphi_{n,n+1}\rangle$,
the inverse Debye-Waller factor calculated in Section 6. The
fluctuating dielectric constant leads to a strong and random
refraction which represents just the necessary transformation
mechanism \cite{ser}.
\par
As for dissipation mechanism two ideas are discussed. The first
\cite{ser} is the dissipation by the normal component which 
exists in SC with the nodes of the OP in magnetic field
even at zero temperature \cite{volovik}: the normal carriers
appear near vortices along special directions. The density
of normal carriers is estimated as $n\sqrt{B/H_{c2}}$. This
mechanism leads to a linewidth $\Delta\omega /\omega\sim 0.1$
in agreement with the experiment. However, it is not clear
whether the growth of the relative width of the resonance line
with decreasing temperature predicted by this mechanism
is compatible with the experimental data \cite{tsui1},
\cite{tsui2}, \cite{matsuda}.
\par
The second dissipation mechanism is via the excitations of the
phase oscillations in superconductors with disordered
vortex arrangement. As it was discussed above the disordered
vortices generate strong static phase fluctuations. Due to
nonlinearity of the Josephson energy, these fluctuations
may create inhomogeneous localized phase oscillations.
They are the same plasma oscillation in an inhomogeneous
situation. One can expect a continuos spectrum of these
oscillations starting with zero energy. The excitation
of these oscillations represents a kind of the inhomogeneous
line broadening. Estimates made by L.N Bulaevsky et al.,
\cite{bul-dis}  give the relative line-width $\approx 0.1$. 
They also predict that the
relative line-width does not depend on the temperature.
\par
In the same preprint, Bulaevsky et al. have also considered 
the well-known Bardeen--Stephen mechanism \cite{Bar-Steph} 
applied to the vortices
oscillating near their equilibrium pinned positions. It should
be taken into account that this dissipation is reduced by
the quantization of quasiparticles in the vortex core. The
quantization is insubstantial for low-temperature SC with
the core size $\xi\sim 1,000\AA$. However, in HTSC $\xi_{ab}
\sim 30\AA$ and the characteristic level spacing is about
$100K$. Since the experiments have been performed starting
from $2K$ the excitation of quasiparticles bound to vortex cores
seems to be irrelevant.

\section{Selectional transparency in magnetic field.}
This is a novel effect which, as we hope, can  be observed at
comparatively weak  magnetic field $B<B_J=\phi_0/
\gm d^2$ parallel-to-layers. It was shown experimentally \cite{zav-zav},
\cite{mansky}, \cite{kadowaki} that the pinning of parallel
vortices is much weaker than the pinning of pancakes.
Therefore, it can be expected that the parallel vortices form
a highly regular lattice even at fields $B<B_J$. As it was
discussed earlier, the PF is not influenced much by the magnetic
field in this region. An incident EM wave is modulated by the
vortex lattice. The modulated wave has the same frequency as the
incident one, but its wave-vector coincides with one of the
vortex lattice reciprocal vectors. A most important situation
is that when the reciprocal vector is parallel to layers. Neglecting
the in-plane anisotropy, an in-plane wave-vector is determined
by an integer $k$ and has the length $q=2\pi k/a$. Thus the
resonance is shifted to a frequency;
\be
\omg =\sqrt{\pff + \alpha v_F^2q^2}
\label{shift}
\ee
where $\alpha$ is a constant of the order of unity. The resonance
frequency can be regulated by a variation of the magnetic field.
If the frequency is fixed, the centers of transparency windows
correspond to following values of fields:
\be
B_k\,=\,\frac{(\omg^2-\pff )\phi_0}{(2\pi v_F\alpha k)^2}
\label{trans-eq}
\ee
The transparency can be observed if the electric vector has
a component perpendicular to the layers. This phenomenon can
be used for the diagnostics of the vortex lattice. For example,
if the vortex lattice is pinned by the periodic potential of
the crystal lattice \cite{feinberg}, \cite{kopnin}, then, instead
of Eq. (\ref{trans-eq}), the resonance frequency obeys a following
equation:
\be
B_k\,=\,\frac{\sqrt{(\omg^2-\pff )\phi_0B_0}}{2\pi v_F\alpha k}
\label{trans-met}
\ee
where $B_0$ is a value of the field at which the vortex lattice
has been quenched. The effect can be used for creating
magnetic field regulated infrared and micro-vawe optical gates.

\section{Conclusions.}
1. In layered structures the PF is strongly reduced, if
electric field is perpendicular to the layers.\\
2. The plasma oscillations in SC can
propagate even if they are overdamped in the normal state.\\
3. The experiments cited in the text convincingly show
the existence of the plasma edge and the resonance plasma
attenuation in three HTSC.\\
4. The plasma oscillations can propagate in rather clean
SC with nodes of the OP, since the Cooper-pair breaking 
is dynamically forbidden.\\
5. In Josephson-linked SC, sufficiently high
magnetic fields reduce the PF dramatically.\\
6. At weaker fields, parallel to the layers, an effect of the
selectional transparency should be observable.\\
7. Measurement of the magnetic field dependence of the PF
can be used for the diagnostics of the vortex state in
HTSC.\\

\begin{center}
{\large\bf Acknowledgements.}
\end{center}
The author is indebted to L.N. Bulaevsky and S.V. Pokrovsky
for cooperation and to T. Mishonov, I. Bozovic and Vl. Kresin
for interesting discussions. Thanks are due to the authors of 
the works \cite{koch}, \cite{tamasaku}, \cite{tsui1} for permission
to use their experimental curves in this review.

\pagebreak
\begin{center}
{\large\bf Figure Captions}
\end{center}
{\bf Fig. 1}.\\
Theoretical curves for the reflectivity vs frequency measured
in units $\pf$ \cite{PP}. The high-frequency dielectric constant $\epi = 25$. 
Different curves correspond to different values of the scattering rate:
$\pf\tau\,=\,0.01;0.1;0.5;;1.0;5.0$.\\
{\bf Fig. 2}.\\ 
The plasma edge in $YBCO$ \cite{koch}. Far infrared reflectivity at 300 K 
(solid line), 110 K (dashed line), 60 K (dotted line), and 10 K (dash-dotted
line). Upper panel: electric field parallel to the $ab$-plane. Lower panel:
electric field parallel to the $c$-axis.\\
{\bf Fig. 3}.\\
The plasma edge in $La_{2-x}Sr_xCuO_4$ \cite{tamasaku}. The electric field
is parallel to the $c$-axis. The values of the concentration $x$ and 
temperature $T$ are indicated on the graphs.\\
{\bf Fig. 4}.\\
The resonance attenuation in $Bi_2Sr_2CaCu_2O_{8+\delta}$ \cite{tsui1}. The 
surface resistance $R_s$ vs magnetic field observed at 30 GHz and 2.8 K.
A remarkable hysteresis is caused by flux trapping in the crystal. In the
inset the geometry of the experiment is shown. 
\end{document}